\input harvmac
\input epsf

\newcount\figno
\figno=0
\def\fig#1#2#3{
\par\begingroup\parindent=0pt\leftskip=1cm\rightskip=1cm\parindent=0pt
\baselineskip=12pt
\global\advance\figno by 1
\midinsert
\epsfxsize=#3
\centerline{\epsfbox{#2}}
\vskip 14pt

{\bf Fig. \the\figno:} #1\par
\endinsert\endgroup\par
}
\def\figlabel#1{\xdef#1{\the\figno}}
\def\encadremath#1{\vbox{\hrule\hbox{\vrule\kern8pt\vbox{\kern8pt
\hbox{$\displaystyle #1$}\kern8pt}
\kern8pt\vrule}\hrule}}

\overfullrule=0pt

\noblackbox
\parskip=1.5mm

\def\Title#1#2{\rightline{#1}\ifx\answ\bigans\nopagenumbers\pageno0
\else\pageno1\vskip.5in\fi \centerline{\titlefont #2}\vskip .3in}

\font\caps=cmcsc10

\noblackbox
\parskip=1.5mm

  
\def\npb#1#2#3{{\it Nucl. Phys.} {\bf B#1} (#2) #3 }
\def\plb#1#2#3{{\it Phys. Lett.} {\bf B#1} (#2) #3 }
\def\prd#1#2#3{{\it Phys. Rev. } {\bf D#1} (#2) #3 }
\def\prl#1#2#3{{\it Phys. Rev. Lett.} {\bf #1} (#2) #3 }
\def\mpla#1#2#3{{\it Mod. Phys. Lett.} {\bf A#1} (#2) #3 }

\def\cmp#1#2#3{{\it Commun. Math. Phys.} {\bf #1} (#2) #3 }

\def\bb#1{{\tt hep-th/#1}}

\def\jhep#1#2#3{{\it J. High Energy Phys.} {\bf #1} (#2) #3 }


           \def\CO{{\cal O}} \def\CZ{{\cal Z}}
   
\def\CL{{\cal L}}   
   
  \def\CK{{\cal K}} \def\CQ{{\cal Q}}


\def\dj{\hbox{d\kern-0.347em \vrule width 0.3em height 1.252ex depth
-1.21ex \kern 0.051em}}

\def\Tr{{\rm Tr\,}}
\def\tr{{\rm tr\,}}

\def\ket{\rangle}
\def\bra{\langle}

\def\ozeta{\overline {\zeta}} 
\def\otheta{\overline{\theta}}

\def\pt{\partial}

\def\Dirac{\,\raise.15ex\hbox{/}\mkern-13.5mu D}
\def\dirac{\,\raise.15ex\hbox{/}\kern-.57em \partial}
\def\shalf{{\ifinner {\textstyle {1 \over 2}}\else {1 \over 2} \fi}} 
\def\sshalf{{\ifinner {\scriptstyle {1 \over 2}}\else {1 \over 2} \fi}} 
\def\sfourth{{\ifinner {\textstyle {1 \over 4}}\else {1 \over 4} \fi}}

\lref\rlargen{G. 't Hooft, \npb{75}{1974}{461.}}

\lref\rmaldacobi{N. Itzhaki, J. Maldacena, J. Sonnenschein and S. Yankielowicz,
\prd{58}{1998}{046004} \bb{9802042.}
}

\lref\rme{J.L.F. Barb\'on,  \plb{543}{2002}{283}  
\bb{0206207.}}  

\lref\rads{E. Witten,
{\it Adv. Theor. Math. Phys.} {\bf 2} (1998)
253 \bb{9802150.} S.S. Gubser, I.R. Klebanov and A.M. Polyakov,
\plb{428}{1998}{105} \bb{9802109.}}

\lref\rexpri{J. Maldacena and A. Strominger, \jhep{9812}{1998}{005}
\bb{9804085.}}

\lref\rHP{G.T. Horowitz and J. Polchinski, \prd{55}{1997}{6189}
\bb{9612146.}}

\lref\rtroost{J. Troost, \bb{0308044.}}

\lref\rkgub{S.S. Gubser and I.R. Klebanov,  
\npb{656}{2003}{23}  
\bb{0212138.}}
 
\lref\rberkooz{M. Berkooz, A. Sever and A. Shomer,  \jhep{0205}{2002}{034} \bb{0112264.}}

\lref\rwitm{E. Witten, \bb{0112258.}}

\lref\rotros{P. Minces and V.O. Rivelles, \jhep{0112}{2001}{010} \bb{0110189.}
W. Muck,
\plb{531}{2002}{301}
\bb{0201100.}
P. Minces,
\prd{68}{2003}{024027} \bb{0201172.}
A. Sever and  A. Shomer,
\jhep{0207}{2002}{027} \bb{0203168.}}

\lref\rgm{S.S. Gubser and I. Mitra, \bb{0009126}. \jhep{0108}{2001}{018}  
\bb{0011127.}    
H. Reall, \prd{64}{2001}{044005} 
\bb{0104071.}}  

\lref\rwitmast{E. Witten, {\it ``Recent Developments in Gauge Theories"}, 1979
Cargese Lectures. Ed. G. 't Hooft et. al. Plenum (1980).}

\lref\rexodef{O. Aharony, M. Berkooz and E. Silverstein, \jhep{0108}{2001}{006}
\bb{0105309.}  \prd{65}{2002}{106007}  \bb{0112178.}}

\lref\rthresholds{J.L.F. Barb\'on, I.I. Kogan and E. Rabinovici,
\npb{544}{1999}{104} \bb{9809033.}}

\lref\rHPage{S.W. Hawking and D. Page, \cmp{87}{1983}{577.}}

\lref\rwitbar{E. Witten, \npb{160}{1979}{57.}}

\lref\rgl{R. Gregory and R. Laflamme, 
\prl{70}{1993}{2837} 
\bb{9301052.}}

\lref\rwitheta{E. Witten, \prl{81}{1998}{2862}
\bb{9807109.}}

\lref\retap{E. Witten, \npb{156}{1979}{269.} G. Veneziano, \npb{159}{1979}{
213.}}

\lref\rwittenold{E. Witten, {\it Ann. Phys.} {\bf 128} (1980) 363.}

\lref\rwitthp{E. Witten, {\it Adv. Theor. Math. Phys.} {\bf 2} (1998)
505 \bb{9803131.}}

\lref\rmalda{J. Maldacena, {\it Adv. Theor. Math. Phys.} {\bf 2} (1998)
231 \bb{9711200.}}

\lref\rgubmit{S.S. Gubser and I. Mitra,  
\prd{67}{2003}{064018}  
\bb{0210093.}}
 
\lref\roldmm{S.R. Das, A. Dhar, A.M. Sengupta and S.R. Wadia, \mpla{5}{1990}{1041.}
L. Alvarez-Gaum\'e, J.L.F. Barb\'on and C. Crnkovic, \npb{394}{1993}{383.} G.
Korchemsky, \mpla{7}{1992}{3081,} \plb{256}{1992}{323.} I.R. Klebanov,
\prd{51}{1995}{1836.} I.R. Klebanov and A. Hashimoto, \npb{434}{1995}{264.}
J.L.F. Barb\'on, K. Demeterfi, I.R. Klebanov and C. Schmidhuber, \npb{440}{1995}{189.}}


\baselineskip=15pt

\line{\hfill CERN-TH/2003-283}
\line{\hfill IFT-UAM/CSIC-03-48}
\line{\hfill FTUAM-03-27}
\line{\hfill {\tt hep-th/0311274}}

\vskip 0.2cm

\Title{\vbox{\baselineskip 12pt\hbox{}
 }}
{\vbox {\centerline{AdS/CFT, Multitrace Deformations and   }
\vskip10pt
\centerline{New Instabilities of Nonlocal String Theories}
}}

\vskip0.3cm

\centerline{$\quad$ {\caps 
 J.L.F. Barb\'on~$^{a,}$\foot{ On leave
from Departamento de F\'{\i}sica de Part\'{\i}culas da 
Universidade de Santiago de Compostela, Spain.} and   
C. Hoyos~$^{b}$ 
}}
\vskip0.3cm

\centerline{{\sl $^a$ Theory Division, CERN,
 CH-1211 Geneva 23, Switzerland}}
\centerline{{\tt
barbon@cern.ch}}

\vskip0.3cm 

\centerline{{\sl $^b$ Instituto de F\'{\i}sica Te\'orica UAM/CSIC,  C-XVI}}  
\centerline{\sl and Departamento de F\'{\i}sica Te\'orica, C-XI } 
\centerline{{\sl Universidad Aut\'onoma de Madrid, E-28049--Madrid, Spain }}
\centerline{{\tt  c.hoyos@uam.es}}

\vskip0.3cm

\centerline{\bf ABSTRACT}

 \vskip 0.3cm

 \noindent 

We study ``multitrace" deformations of large $N$ master fields in models with
a mass gap. In particular, we determine the conditions for the multitrace couplings
to drive tachyonic instabilities. These tachyons  represent new local
instabilities of the associated nonlocal string theories. In the particular case
of D$p$-branes at finite temperature, we consider topology-changing phase transitions
and the effect of multitrace perturbations on the corresponding phase diagrams.

\vskip 1.5cm

\Date{November 2003} 
               

\vfill





\baselineskip=15pt

\newsec{Introduction}

\noindent

The AdS/CFT correspondence \refs{\rmalda, \rads}  provides a nonperturbative
definition of quantum gravity  in asymptotically Anti-de Sitter (AdS) 
  spaces,
in terms of a quantum field theory with an ultraviolet fixed point (CFT), 
 formally living on a conformal boundary of the AdS space. 
The prototype examples define string theory on certain
 ${\rm AdS}_{d+1}$ vacua as dual 
to supersymmetric
$SU(N)$ 
Yang--Mills (SYM) theories in the large $N$ limit of 't Hooft \refs\rlargen,
defined on ${\bf R}\times {\bf S}^d$.  
In this correspondence,
 string perturbation theory is obtained as the $1/N$ expansion
of the gauge theory, and different semiclassical backgrounds with a common
AdS asymptotics correspond to different ``master fields" of the gauge theory
\refs\rwitmast.  

Single-string states in perturbation theory arise as single-trace states
in the Yang--Mills theory, such as $W_\gamma \,|{\rm vac}\,\ket$, where   
\eqn\wils{
W_\gamma = \tr\;{\rm P} \;\exp\,\left(i\oint_\gamma A\right) 
}
for some contour $\gamma$. For CFT's one has an exact operator-state mapping,
so that we may equivalently talk about the space of local operators of the CFT
on ${\bf R}^d$.  
For finite values of $N$, i.e. at the nonperturbative
level, this correspondence between single strings and single traces
 must break down, since the  Yang--Mills theory with
gauge group of rank $N$  has
only $O(N)$ independent single-trace {\it local} operators. Many 
single-trace local operators are linearly related to
 multitrace local operators, which means that the perturbative Fock space
of string perturbation theory loses its exact meaning for excitations with
 $O(N)$ strings. This is related to the so-called ``stringy exclusion
principle" \refs\rexpri. 

One way of studying nonperturbative effects associated with these phenomena
is to consider deformations of the AdS/CFT correspondence by  
``condensates" of multistring states in the AdS space, 
i.e. deformations of the CFT by multitrace operators. Such backgrounds were
studied in \refs\rexodef\ and argued to provide new classes of string
theories with nonlocal interactions, both on the  worldsheet and the
target space  (NLST). 
A nontrivial generalization of the AdS/CFT correspondence
is needed to describe these deformations in the leading large $N$ approximation
\refs{\rberkooz, \rwitm, \rotros}. 

It was shown in \refs{\rwitm, \rme}\ that these AdS/CFT prescriptions are equivalent
to the Hartree approximation from the point of view of the gauge theory,
in the spirit of \refs\rwitbar.  
 The basic gauge invariants are single-trace normalized operators
of the form $\CO_n =  N^{-1} \,\tr\,F^n$. Interactions that are non-linear
on the basic invariants,  such as  
\eqn\mint{
\CL_{\rm int} \sim c_m \,(\CO_n)^m\;,
}
 may be substituted by
a linear interaction 
\eqn\har{
\CL_{\rm eff} \sim C_n\,\CO_n
\,,}
with $C_n$ a ``collective potential" that is determined selfconsistently. 
The basic identity determining $C_n$ is the saddle-point equation at
$N=\infty$. 
The expectation value of the field equation for $\CO_n$
 gets a contribution from
\mint\ of the form
\eqn\see{
m\,c_m \,\bra (\CO_n)^{m-1} \ket 
\longrightarrow m\,c_m \,(\bra \CO_n \ket)^{m-1}\;,}
where we have used the large $N$ factorization  of gauge-invariant operators
in the right hand side. Since $\bra \,\CO_n \,\ket$ are pure numbers, 
this is the contribution of a linear action \har\
to the field equation, provided we adjust the ``effective potential" 
according to
\eqn\seee{
C_n = m\,c_m\,(\bra \CO_n \ket)^{m-1} \;.}
In the linear model \har\ we may calculate the expectation values $\bra \CO_n \ket$
by taking derivatives with respect to $C_n$, so that  the expectation values
are functions of $C_n$, and \seee\  becomes an equation
that determines $C_n$ selfconsistently.

More generally, for an arbitrary action functional with 't Hooft's scaling    
\eqn\ac{
S= N^2 \,\int d^d x\; L(\,\CO_1, \CO_2, \dots)\;,}
we can introduce an effective single-trace model
\eqn\ach{
{\overline S} = N^2 \,\sum_n \int d^d x\;{\overline \zeta}_n \;\CO_n \;, }  
where the effective couplings $\ozeta$
 are determined by  the  solution of the ``master equations":
\eqn\meq{
{\overline \zeta}_n = {\pt L \over \pt \CO_n} \left(\bra \CO \ket_{\overline 
\zeta}\right)\;.}
When interpreted in the context of the AdS/CFT correspondence, \meq\
is equivalent to the boundary conditions proposed in \refs\rads.

It was pointed out in \refs\rme\ that \meq\ can have various solutions
for each of the effective couplings $\zeta_n$. Hence, the ``master field"
of the theory with multitrace interactions can develop a branched structure.
On the dominating branch the partition function is maximized or,   
equivalently,  the
large $N$ vacuum energy is minimized. When level-crossing phenomena cause
the dominating branch to change, we will have in general large $N$ phase
transitions. These phase transitions correspond to nonperturbative 
tunneling processes between different backgrounds of the NLST. Similar phenomena
were extensively studied in the context of $c\leq 1$ matrix models \refs\roldmm.   

In this paper we extend the results of \refs\rme\ on particular  
solutions of \meq\ for specific examples.  
In addition, we study more carefully the structure of large $N$
phase transitions induced by multitrace perturbations, and we compare  
their effects with other types of large $N$ phase transitions  that can
be calculated using the AdS/CFT correspondence. We study the conditions
for the NLST backgrounds to be locally stable, and we find a new kind
of tachyonic instabilities triggered by the multitrace interactions.   

The article is organized as follows. In section 2 we discuss the general
structure of the $1/N$ expansion around a background with multitrace
interactions. In section 3 we  study the local stability of the 
saddle points, including  the new tachyonic instabilities
of multitrace origin and a detailed discussion of one concrete example.
  In section 4 we study multitrace effects on one particular instance of
{\it global} instabilities, namely  topology-changing transitions
on toroidal backgrounds, generalizations of the Hawking--Page transition
in the pure AdS case \refs{\rHPage, \rwitthp}.  We end with the conclusions
and  an appendix dealing with  subtleties arising in the case of
the topological-charge operator in four-dimensional Yang--Mills theories.          

\newsec{Systematics of the Mean Field Approximation}

\noindent

Let us consider a 
 gauge-theory model specified by a single-trace action $S_0$ and 
a multitrace perturbation by a general function  of a single-trace
operator of the form 
\eqn\oper{
\CO \sim {1\over N} \,\tr\,F^n + \dots
}
where the dots stand for other terms in gauge or matter fields.  The
complete action is 
\eqn\defa{
S = S_0 +  N^2\int d^d x \;\zeta\;\CO + N^2 \int d^d x \; f(\CO)\;,}  
where we have separated explicitly the linear part of $f(\CO)$. We shall assume
that operator condensates in this theory, $\bra\,\CO\,\ket$, are determined
 in terms of the microscopic couplings, with at most a discrete degeneracy.
In particular, this means that we will only consider theories whose
single-trace limit $f\rightarrow 0$ has  isolated vacua with a mass gap,
separated from any possible moduli spaces of vacua.  

The coupling $\zeta$ will be promoted to a source for the operator $\CO$, although
the large $N$ master field will be assumed translationally invariant on
${\bf R}^d$, and $\zeta$ will be evaluated as a constant in all expectation values.  

In the Hartree approximation, the expectation value $\bra \CO \ket$ may
be calculated at $N=\infty$   from the single-trace effective theory
with action
\eqn\effe{
{\overline S} = S_0 + N^2 \int d^d x  \;{\overline \zeta}\;\CO\;,}
with ${\overline \zeta}$ determined from  the master equation: 
\eqn\meqq{
{\overline \zeta} = \zeta + f' \left(\, \bra \CO \ket_{\overline \zeta}
\,\right)\;.}

The consideration of large $N$ phase transitions requires comparing
the values of the full  partition function
\eqn\fullp{
\CZ\,[\,\zeta, f\,] = \int DA\;e^{-S} }
 at different large $N$ master
fields. The path integral measure over the gauge field $A$ 
 may contain various other fields as well, although we use the notation $DA$ for
simplicity.
  The effective single-trace partition function 
\eqn\eep{
{\overline \CZ} \,\left[\;{\overline \zeta}\;\right] = \int DA \;e^{-{\overline S}} }
can be used to compute the single-trace expectation values, but it
is in general different from \fullp\ at the saddle points.\foot{This wrong
assumption was made in \refs\rme\ in a brief analysis of some examples.
In the present work we correct these errors, although they do not change
the qualitative picture.}      

In order to derive a useful expression for $\CZ[\zeta, f]$ it is convenient
to introduce appropriate auxiliary fields. First, we insert a delta-functional
constraint by the identity
\eqn\idel{
{\bf 1} = \int \prod_x d\sigma_x  \;\delta \,[ \sigma - \CO ]\;,}
which defines $\sigma(x)$ as a classical interpolating field for the local operator 
$\CO(x)$. We can further exponentiate the delta-functional by means of a second
auxiliary field  
\eqn\delf{
\delta \,[\sigma -\CO ] = \int \prod_x {d\chi_x \over 2\pi} \;e^{i\int \chi (
\sigma -\CO)} \;.}
It will be convenient to distribute democratically the factor of $(2\pi)^{-1}$
between the $\sigma$ and $\chi$ measures. Defining
$$
D \sigma = \prod_x {d\sigma_x \over \sqrt{2\pi}}
$$
and analogously for $D \chi$, we can write
\eqn\iddel{
{\bf 1} = \int D \sigma \; D \chi \;e^{i\int \chi (\sigma - \CO)} 
\;, }
which defines a formal path integral representation of the full partition
function, 
\eqn\fpi{
\CZ[\zeta, f] = \int D\sigma \,D\chi\;\CZ_0 \,\left[\zeta + i{\chi \over N^2}
 \right] \;\exp\,\left[-N^2 \int f(\sigma) + i\int \chi\,\sigma \right]\;.}
In this expression $\CZ_0$ denotes the single-trace partition function that
results by setting $f=0$.  
The  connected functionals of the full and single-trace theory 
are     given by 
\eqn\stef{
W [\,\zeta, f\,] = -{1\over N^2} \;\log\;\CZ \,[\, \zeta, f\,] \;\;, \qquad 
W_0 \,[\zeta ] = -{1\over N^2} \;\log\;\CZ_0 \,[ \zeta] \;. }
In these definitions, $\zeta$ is treated as a spacetime-dependent source. At the
saddle points we will assume translational invariace on ${\bf R}^d$ and it
 will be useful to define  the volume densities 
\eqn\dens{
W [\,\zeta, f\,] = \int d^d x \; w (\zeta, f)  \;\;,\qquad
 W_0 \,[\zeta ] = \int d^d x \; w_0 (\zeta) \,,}
as a function of the microscopic  couplings. 

 Both $W_0$ and all its functional derivatives have a $1/N^2$ expansion
with leading term of $O(1)$. They generate the set of connected correlators
of $\CO (x)$ in the single-trace theory.
 For example, the one-point function is of $O(1)$,
\eqn\onep{
\bra\,\CO \,\ket_\zeta = {\delta W_0 \over \delta \zeta} = w'_0 (\zeta)\;,}
while the connected two-point function is of  $O(1/N^2)$,
\eqn\twopp{
\bra\,\CO (x_1) \;\CO(x_2) \,\ket_{c,\;\zeta} = -{1\over N^2} \,{\delta W_0 \over
\delta \zeta(x_1) \,\delta \zeta(x_2)} \;.}

  In order to derive a $1/N$ expansion for $\CZ\,[\zeta, f]$ we evaluate 
the path integral over the auxiliary fields in the saddle-point approximation.
We write $\sigma = \sigma_c + \sigma' /N$ and $\chi = \chi_c + N \chi'$, and
we determine $\sigma_c$ and $\chi_c$ requiring the cancellation of
the $O(N)$ terms. This leads to the equations
\eqn\sadd{
\sigma_c = {\delta W_0 \over \delta \zeta} \Big|_{\zeta + i\chi_c /N^2}
\;,  \qquad    
i\chi_c = N^2 \,f' (\sigma_c) \;.} 
Defining  an effective coupling
\eqn\ebar{
\ozeta = \zeta + {i\chi_c \over N^2} }
we see that
 $\sigma_c = \bra\,\CO \,\ket_{\ozeta}\;$,  and the saddle-point equations
 \sadd\ are equivalent to the master equation \meqq. These equations imply
that the saddle point of the $\chi$ integral lies in the  imaginary
axis for real values of the one-point function. In this case, one must deform
the contour of integration of the zero mode of $\chi$ accordingly. 

The leading term in the $1/N$ expansion of the partition function is
then given by
\eqn\ledd{
w(\zeta, f) = w_0 \,(\,\ozeta\,)
 +  f\left(
\bra \, \CO\,\ket_{\ozeta} \right) -  \bra\,\CO \,\ket_{\ozeta} \;f'
\left(\bra\,\CO\,\ket_{\ozeta} \right) + O(1/N^2)\;.}
This shows that the partition function at large $N$ is not just the
partition function of the effective single-trace model; there are two
extra terms that were omitted in \refs\rme.

The local stability of a given master field is controlled by the functional
quadratic in the perturbation fields $\sigma', \chi'$, of $O(1)$ in the
large $N$ expansion around the saddle point. 
This term contributes
\eqn\stab{
-\log\;\CZ\,\left[\zeta, f\right] = O(N^2) + 
\shalf \,\log\;{\rm Det}\;\left[\,\CK\,\right] +
O(1/N^2)\;,}
where $\CK$ is a nonlocal operator acting on the field-space $(\chi', \sigma')$ 
as 
\eqn\matt{
\CK = \pmatrix{{\overline G}_2  & -i \cr
-i & f_c''  \cr}\;,}
with $f''_c = f'' (\sigma_c)$ and  
\eqn\gdos{
{\overline G}_2 (x_1, x_2) = 
-{\delta^2 \;W_0 \over \delta \,\zeta(x_1)\;\delta\,
\zeta(x_2)}\;\Big|_{\ozeta} = N^2 \;\bra\,\CO(x_1) \,\CO(x_2) \,\ket_{c, \;\ozeta} \;, }
 the connected two-point function in the effective single-trace model.

The higher-order functional derivatives of $W_0$, together with the 
higher derivatives of the multitrace potential $f(\sigma)$, define
effective vertices for the $1/N$ expansion of 
 \fpi. The connected single-trace correlators define nonlocal vertices of the $\chi'$
field, 
\eqn\vww{
V_W = \sum_{n\geq 3} {1\over N^{n-2} } \;{i^n \over n!} \int d^d x_1 \cdots
d^d x_n \; {\delta^n \,W_0 \over
\delta \zeta(x_1) \cdots \delta \zeta(x_n)}
 \Big|_{\ozeta} \;\;\chi'(x_1) \cdots \chi'(x_n) \;,} 
where each functional derivative of $W_0$ has a separate expansion in powers of
$1/N^2$ with leading term of $O(1)$.

Both the saddle-point equations \sadd\ and the propagator \matt\ depend implicitly
on the full single-trace connected functional $W_0$, which itself has a $1/N^2$
expansion with leading term of $O(1)$. In principle, we have the choice of keeping
this implicit $1/N$ expansion in the value of the saddle point $\sigma_c, \,\chi_c$ and
the effective propagator $\CK$. However, we often ignore the explicit form of $W_0$
beyond the planar approximation and, in practice, we solve for $\sigma_c, \,\chi_c$
just at the leading (planar) order. In this case we 
 may include the nonplanar dependence of \sadd\
and \matt\ 
via a series of explicit tadpole and mass insertions. 
 These new vertices 
have the form \vww\ with $n=1, 2$ and $W_0$ replaced by its ``non-planar" part with
leading scaling of $O(1/N^2)$. This means that the tadpoles form a series with leading
term of $O(1/N)$, whereas the mass insertions start at $O(1/N^2)$. 
 
Finally, the vertices for the $\sigma'$ field are local, 
\eqn\sigmp{
V_\sigma = \sum_{m\geq 3} {1 \over N^{m-2}} \;{1\over m!} \int d^d x
 f^{(m)} (\sigma_c)\;
(\sigma'_x)^m\;.}
Using $\CK$ as a propagator and \vww, \sigmp\ as vertices, we can 
calculate  the $1/N$ corrections to the master
 field in a systematic diagram technique.  
For quadratic perturbations, $f''' =0$, the $\sigma$ field is free and can be
explicitly integrated out, leaving only the diagram technique of the $\chi$ field,
as in the treatment of \refs\rkgub. 

Regarding the NLST interpretation, if the single-trace model is associated to
some string background $X_0$ via AdS/CFT or a deformation of it thereof, the
perturbation expansion in powers of the multitrace vertices $f^{(m)}(\sigma_c)$
defines the nonlocal worlsheet interactions, according to \refs\rexodef. The
content of the Hartree approximation is simply that one-point functions and
partition functions may be calculated at large $N$ by working in a modified
single-trace background ${\overline X}$, characterized by effective single-trace
couplings $\ozeta$. However, the physics of the NLST goes much beyond the
one-point functions and the large $N$ vacuum energy. In particular, the stability
of the NLST cannot be inferred directly from the stability of ${\overline X}$ in
the single-trace theory, but instead requires a specific analysis.   

\newsec{Stability of the Master Field}

\noindent

The global stability properties of the perturbed model depend to a large extent on
the global properties of the function $f(\sigma)$ for large values of $\sigma$. If
this function is unbounded from below we can expect a globally unstable model. In
this section we shall concetrate on the local stability properties of a given
saddle point, characterized by a solution $\sigma_c, \,\chi_c$ of the master
equations \sadd.

In momentum space the operator $\CK$ has the block form \matt\ for  each value of
the  momentum. The naive stability conditions demand positivity of the eigenvalues
\eqn\eigv{
\lambda_k^\pm=\shalf \left(f''_c+{\overline G}_2 (k)\right) \pm \shalf  
\sqrt{\left((f''_c-{\overline G}_2(k)\right)^2-4}
\;.}
This amounts to the reality condition
\eqn\eigre{
|f''_c-{\overline G}_2(k)|>2
\;,}
together with the positivity conditions  
\eqn\posit{
f''_c+{\overline G}_2(k)>0
\; \qquad 
{\rm and} \qquad
1+f''_c\;{\overline G}_2(k)>0
\;.}
In ordinary field theories we expect ${\overline G}_2 (k) >0$ for
Hermitian $\CO(x)$, so that the violation of the stability conditions
can only occur for sufficiently  negative values of $f''_c$.
  When these conditions are not met, we can still define the integrals
by analytic continuation in $f''_c$ or, equivalently, by an appropriate contour 
rotation of the  $\sigma', \chi'$ integrals around the saddle point.  
However, in this process  physical quantities  will pick up complex
phases that change their physical interpretation. One simple example of this
phenomenon is the imaginary part of the vacuum energy, which should be interpreted
as the total decay width of the unstable saddle point.  

Since   $\chi$ is a formal auxiliary field, it is not obvious that
all the conditions $\lambda_k^\pm >0$ have the same physical status. For
example, the contribution to the vacuum energy coming from each momentum mode
is given by
\eqn\vaceno{
\shalf \;\log\;(\lambda_k^+ \lambda_k^-) = \shalf \;\log\;\left[
1+ f''_c \,{\overline G}_2 (k)\right] 
\;,}
and simply demanding that 
this contribution  be real imposes the less restrictive
condition
\eqn\stab{
1+ f''_c \,{\overline G}_2 (k) >0\;.}
In fact, to the extent that we are interested in the exact
correlation functions of
the operator $\CO(x)$,  we must focus on the propagation properties of
the $\sigma(x)$ field after $\chi(x)$ has been integrated out, since we have
\eqn\iden{
\bra\, \CO(x_1) \cdots \CO(x_n)\,\ket = \bra\,\sigma(x_1) \cdots \sigma(x_n)\,
\ket
}
as an exact statement in the complete theory.  
Therefore, we first  integrate  over $\chi'$ and obtain an effective action
for $\sigma'$ of the form
\eqn\gamm{
\exp\left(-\Gamma_{\rm eff} [\sigma']\right) =
 {\rm Det}^{-1/2}\;\left[\,{\overline G}_2 \;\right]
\;\exp\left[-\shalf \int \sigma' \,(\;
{\overline G}_2^{\;-1} + f''_c \,) \;\sigma' +
O(1/N)\right]\;,} 
where we have omitted the interaction terms that are suppressed by powers of
$1/N$. Integrating now over $\sigma'$ in the gaussian approximation we obtain
the previous result for the total determinant \vaceno, 
\eqn\pdet{
{\rm Det}^{-1/2} \;\left[\,{\overline G}_2 \;\right]
 \,\cdot\,{\rm Det}^{-1/2} \; \left[\,{\overline G}_2^{\,-1} + f''_c \;\right]
={\rm Det}^{-1/2} \; \left[\,1+f''_c \;{\overline G}_2 \;\right]
\;,} 
where we assume a convenient regularization procedure to make sense of these
manipulations. 
More generally, the $1/N$ expansion of the
 correlators  \iden\ can be calculated
using Feynman rules with effective kinetic term
 \eqn\ksigma{
\CK_\sigma = {\overline G}_2^{\;-1} + f''_c\;.   
}
Hence, the physical stability condition demands positivity of this (Euclidean)
kinetic term for all momenta. Assuming ${\overline G}_2 (p) >0$, 
this condition is equivalent to \stab\ above. 

In order to give a more physical characterization of the stability conditions 
let us consider  the spectral representation of the single-trace two-point function
\eqn\specc{
{\overline G}_2 (p) = \int dz {\rho(z) \over z+ p^2}\;,}
with $\rho(z)>0$ the spectral density (for simplicity, we assume that $\CO(x)$ 
is  scalar and Hermitian). Expanding ${\overline G}_2^{\,-1}$ near $p^2 =0$ we have
\eqn\expi{
{\overline G}_2 (p) \approx {Z_0(\CO) \over p^2 + M_0^2} \;,
}
where the wave-function rescaling, $Z_0(\CO)$, and  the mass gap, $M_0^2$, of the
single-trace theory are assumed to be positive,
\eqn\defss{
Z_0 (\CO) = {\left(\int dz \,\rho(z) /z\right)^2 \over \int dz \,\rho(z)/z^2}\;,\qquad 
M_0^2 = {\int dz\,\rho(z)/z \over \int dz \,\rho(z)/ z^2}\;.}
Hence, the physical stability condition boils down to  
\eqn\stabt{
1+ f''_c \;{\overline G}_2 (0) >0\;.} 
The  single-trace mass gap gets renormalized
and this defines an effective mass gap,
\eqn\gappo{
M_0^2 \longrightarrow M_{\rm eff}^2 = M_0^2 + Z_0 (\CO) \;f''_c \geq 0\;,}
that must be positive for stability.

\subsec{Stability and the Master Equation}

\noindent

We can relate  \stabt\ to the master equation by  evaluating  
 ${\overline G}_2 (0)$. The integrated two-point function at zero
momentum can be formally written as
\eqn\tzero{
\int d^d x \;
{\overline G}_2 (0) = N^2 \,\int d^d x \; \int d^d y\;\bra\,\CO(x) \,\CO(y)
\,\ket_{c, \;\ozeta} = 
-\int d^d x \;w_0'' (\,\ozeta\,)\;.}
 Hence,  the stability condition \stabt\ becomes 
\eqn\stavc{
1-f''_c  \;\;w_0''\,(\,\ozeta\,) >0
\;.}

On the other hand, 
the master equation can be written as $\zeta = H(\,\ozeta\,)$, where 
\eqn\defh{
H(\,\ozeta\,) = \ozeta - f' \left(\,w'_0 (\,\ozeta\,)\,\right)\;.}
This function satisfies 
\eqn\stab{
H' (\,\ozeta\,) = 1-f''_c \;\;w''_0 (\,\ozeta\,)\;,}
which is the expression appearing in the
stability condition \stavc. Therefore, $H(\,\ozeta\,)$ is
monotonically increasing (decreasing) for stable (unstable) solutions of
the master equation. 
We can thus determine the stability 
of the solutions by a simple glance at the plot of the function
$H(\,\ozeta\,)$ (c.f. Fig 1). In particular, the solutions of
$H'(\,\ozeta\,)=0$ mark the stability boundary of a given branch
and correspond to the onset of the nonlocal tachyons.

\fig{\sl Sketch of a  function $H(\,\ozeta\,)$ leading to
various solutions of the master equation. Stable branches are
indicated in solid lines, separated by locally unstable branches
in dotted lines. At the extrema of $H(\,\ozeta\,)$ we have
the onset of the tachyonic instabilities.
}{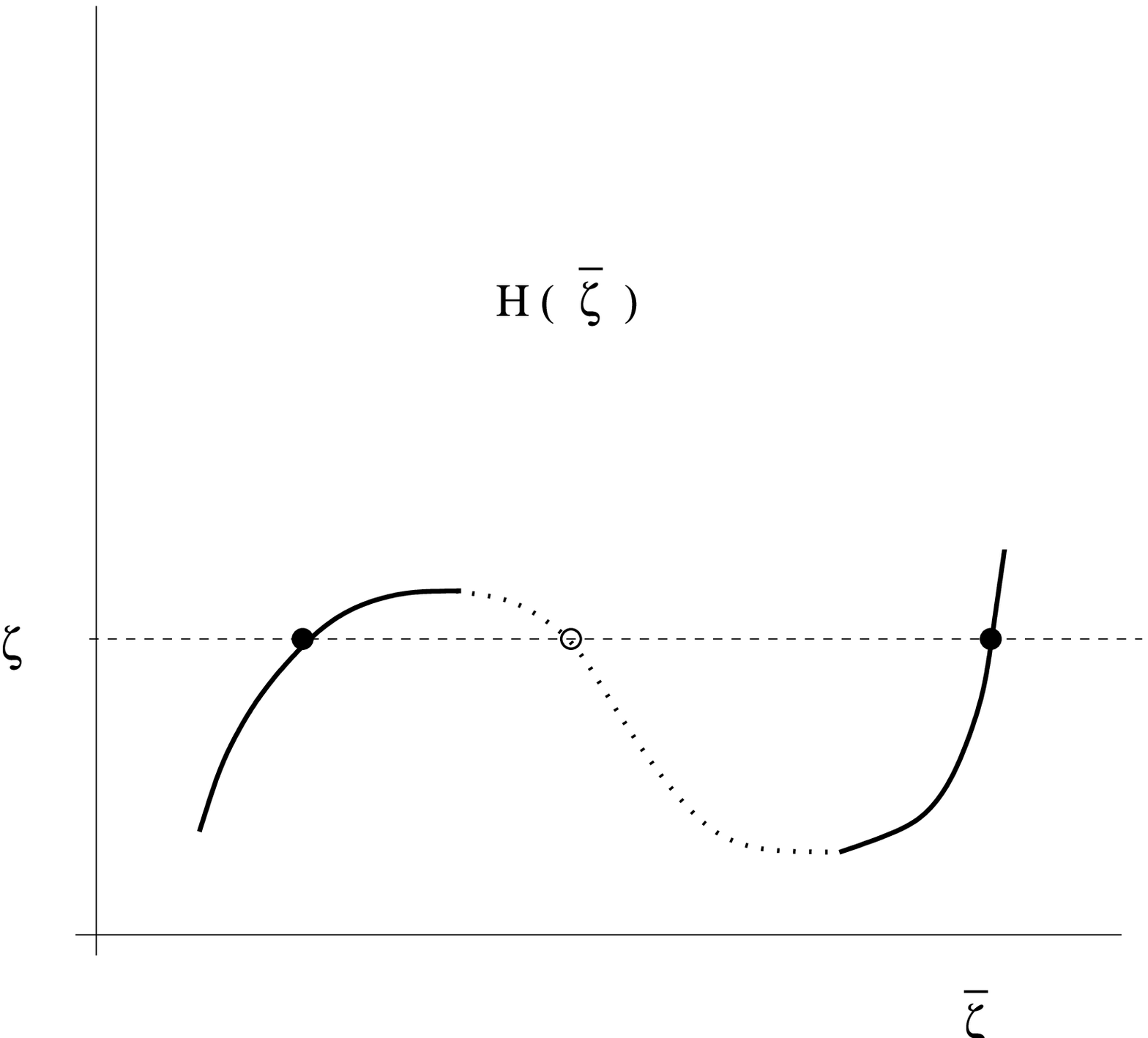}{3truein}

The function $H(\,\ozeta\,)$ also controls
the monotonicity of the vacuum energy with respect to the effective coupling
$\ozeta$. Starting from the general expression \ledd\ we can write the 
vacuum energy density as
\eqn\reden{
w = w_0 + f(w'_0) - w'_0 \,f'(w'_0)\;,}
where all functions depend implicitly on $\ozeta$. Taking the derivative with
respect to $\ozeta$ and using the definition of $H(\,\ozeta\,)$ in \defh\ we
find
\eqn\monow{
w'(\,\ozeta\,) = w'_0 (\,\ozeta\,) \,H'(\,\ozeta\,)\;.}
If the condensate of the single-trace theory, $\bra \CO\ket \sim w'_0$, is
a monotonic function of $\ozeta$ there is a correlation between
the monotonicity of $H(\,\ozeta\,)$ and that of the total vacuum energy.  

It would be interesting to provide a more geometrical interpretation
of these instabilities from the point of view of the NLST. 
The worldsheet interpretation of the multitrace insertions involves contact
interactions between smooth worldsheets \refs{\roldmm, \rexodef}.  
The tachyonic modes described here should be associated to the condensation
of these contact interactions, producing degenerating worldsheets. It would
be interesting to elucidate the geometrical interpretation of these
tachyons in spacetime. See 
\refs\rtroost\ for recent results in this direction.

\newsec{An Explicit Example}

\noindent

A simple model realizing these phenomena is the standard  
``approximation" of QCD in terms of D$p$-branes (with $p<5$)
 at finite temperature
\refs\rwitthp. One starts by engineering a non-supersymmetric version
of Yang--Mills theory in $d=p$ Euclidean dimensions 
by considering the low-energy limit of
a hot D$p$-brane. 

The important expansion parameter is the effective 't Hooft coupling
of the ${\rm YM}_d$ theory at the energy scale  set by the temperature
$T$ of the ${\rm SYM}_{d+1}$ theory on the hot D-brane (c.f. \refs{
\rmaldacobi, \rthresholds}),
\eqn\cueff{
\lambda_d \sim g_s\,N\,\left(\sqrt{\alpha'}\right)^{d-3} \,T^{\;d-3}\;,}
where $g_s$ is the string coupling and $\alpha'$ the string's Regge
slope.  For $\lambda_d \ll 1$ perturbative Feynman diagrams give
a good description, whereas for $\lambda_d \gg 1$ we can use the
AdS/CFT dual in terms of the near-horizon metric of the black D-branes
\refs\rmaldacobi.

The single-trace action is given by
\eqn\sta{
S_0 = {N^2 \over \lambda_d} \int d^d x \;\CL_d \;, \qquad
\CL_d = {T^{\;d-4} \over 2N} \;\tr \,F^2 + \dots\;,}
where the dots stand for regularization artefacts at the temperature
scale $T$ or above 
 (superpartners, higher-dimensional modes or string excitations).  
In what follows we shall simplify the notation by adopting  
 units in which $T=1$.

At $\lambda_d \gg 1$ the supergravity approximation yields an explicit
value for the vacuum energy in terms of the free energy of the hot
D-branes,
\eqn\fen{
w_0 = -(5-d) \,C_d \, \lambda_d^{\;{d-3 \over 5-d}}\;,}
where $C_d$ is a positive constant. From here one finds
\eqn\derivs{
 \bra\,\CL_d \,\ket = w'_0 =  (d-3)\,C_d \,\lambda_d^{\;{2\over 5-d}}\;,
\qquad   
 G_2 (0) = -w''_0 = {2(d-3) \over 5-d} \;C_d \,\lambda_d^{\;
{7-d \over 5-d}}\;.}
Hence, we have all the ingredients needed to consider multitrace
deformations by a nonlinear function of the operator $\CL_d (x)$, 
\eqn\defff{
S=S_0 + N^2 \int d^d x \,f(\,\CL_d\,)\;.}
Defining 
\eqn\efss{
\zeta = {1\over \lambda_d} \;, \qquad H(\,\ozeta\,) = \ozeta-f' \left(
(d-3)C_d \,\ozeta^{\;{2\over d-5}}\right)\;,}
we have an effective single-trace model determined by the inverse 't Hooft coupling $\ozeta$, 
which acts as a curvature expansion parameter of the black 
D$p$-brane metric. From this model we can calculate the condensate
$\bra\, \CL_d\,\ket$ as a function of $\ozeta$, which in turn is determined by 
the master equation $\zeta = H(\,\ozeta\,)$.  

The multitrace deformation is trivial
for $d=3$ in agreement with the fact that the D3-brane free energy
is independent of the dilaton in the leading supergravity approximation.
Incidentally, we notice that $G_2 (0) = - w''_0 <0$ for $d<3$. This
violation of the positivity of the two-point function at zero momentum
is presumably due 
to the  non-Hermiticity of the effective Lagrangian operator \sta, which
would be dominated at $\lambda_d \gg 1$ 
by the regularization artefacts. This
fact renders the $d<3$  models rather unphysical for the matters
discussed here. Therefore, in the following we restrict attention to
$d=4$ and drop the $d$-dimensional subscript from all quantities. We
also simplify the formulas by setting  $C_4 =1$ with an appropriate
choice of coupling parameters.         

The transition between the perturbative and supergravity descriptions
occurs at the ``correspondence line" of \refs\rHP. For the single-trace
model it is given by $\zeta \sim 1$, which is  perturbed  by multitraces
to $\ozeta \sim 1$ or, in terms of the original coupling
\eqn\corro{
\zeta = H(1) = 1-f'(1)\;.}
We see that, depending on the sign of $f'(1)$, the multitraces increase
or decrease the supergravity domain in $\zeta$-space.    
We now discuss some specific choices for the multitrace perturbation.

\fig{\sl The function $H(\,\ozeta\,)$ for the monomium perturbation
with $n=1$ and $\xi= 0.1$ (full line), $\xi=-0.1$ (dashed line).}{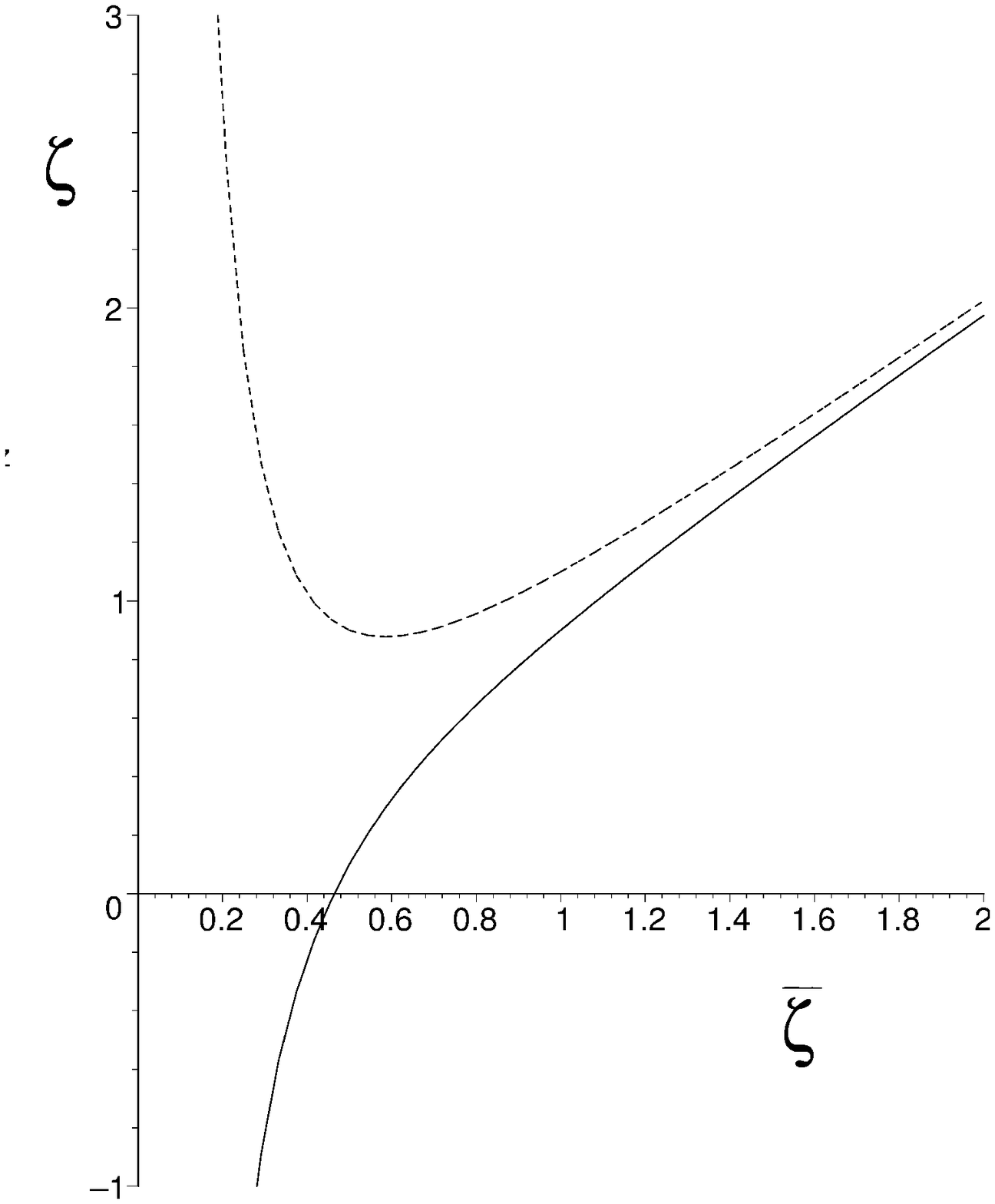}{3truein}

\subsec{Monomium Perturbation}

\noindent

Choosing a multitrace deformation by a single positive power
\eqn\spp{
f(z)= \xi \;  \;{z^{n+1} \over n+1} \;, }
we have
\eqn\heq{
H(\,\ozeta\,) = \ozeta - {\xi \over \ozeta^{\;2n}}\;}
and a vacuum energy density
\eqn\vaee{
w=-\ozeta^{\;-1} - \xi \,{n \over n+1} \,\left(\,\ozeta^{\;-2}\,\right)^{n+1}\;.}

\fig{\sl Binomium perturbation
 $n=2$, $l=1$, $\xi=-0.1$, $\eta=1,0.6,0.01$ (the larger $\eta$ corresponding
to the full line).}{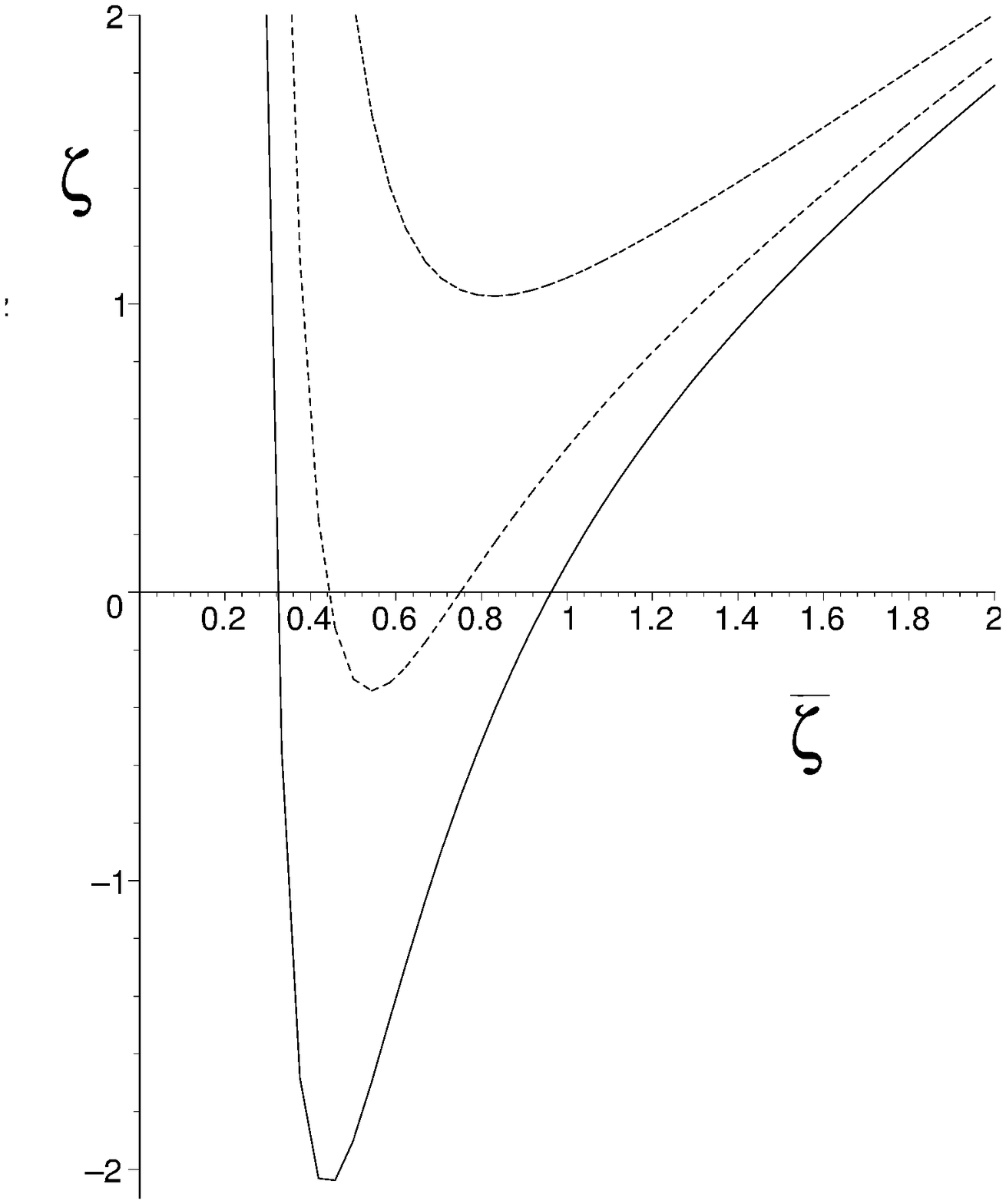}{3truein}

\fig{\sl Binomium perturbation
  $n=2$, $l=1$, $\xi=0.1$, $\eta=-1,-0.6,-0.01$. (the larger $\eta$ corresponding
to the full line)}{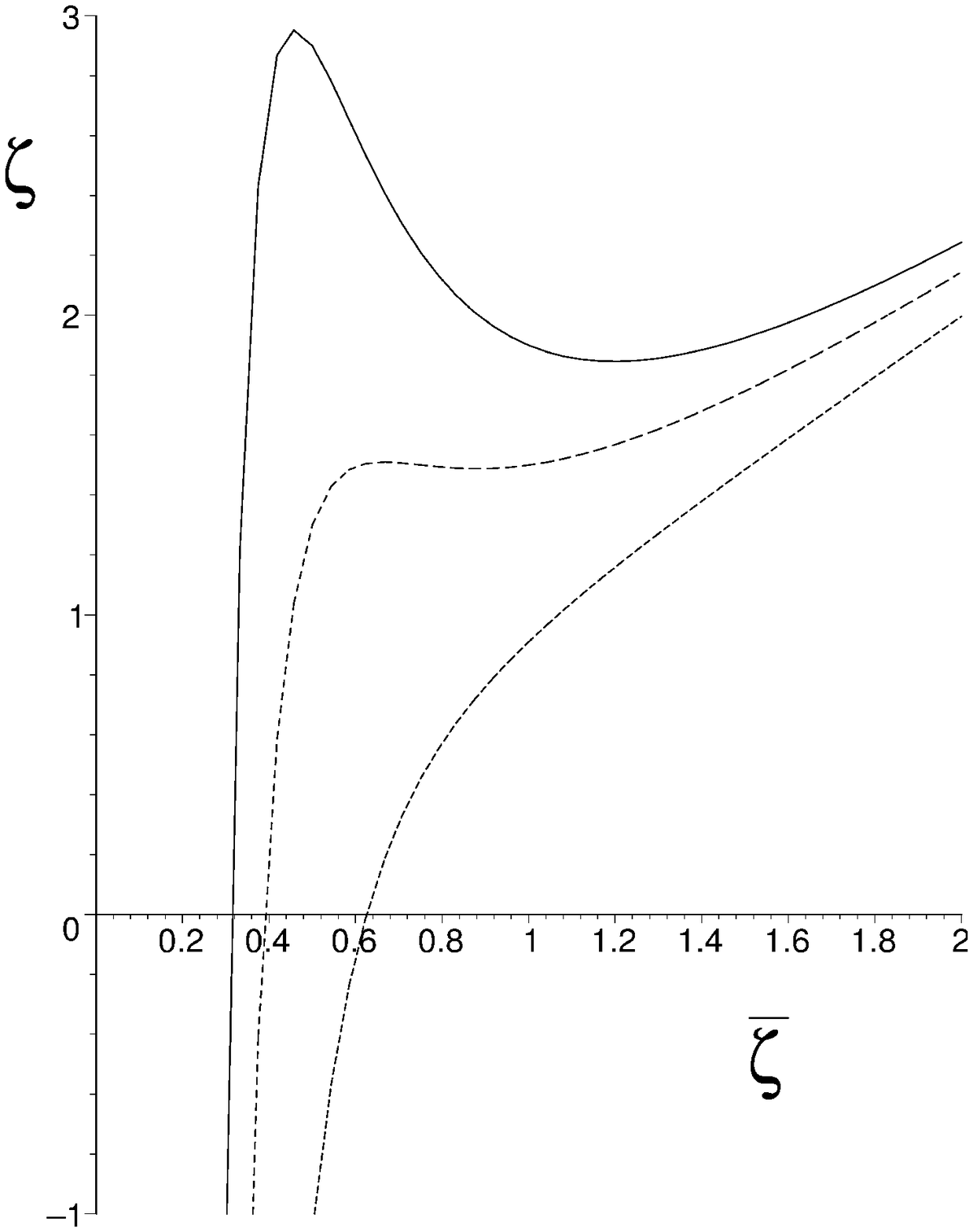}{3truein}

\fig{\sl Logarithmic perturbation with $\varepsilon =1$, $\xi=0.1$ (dashed lines) and
$\xi =-0.1$ (full lines). The branches at $\ozeta >1$ are unphysical.}{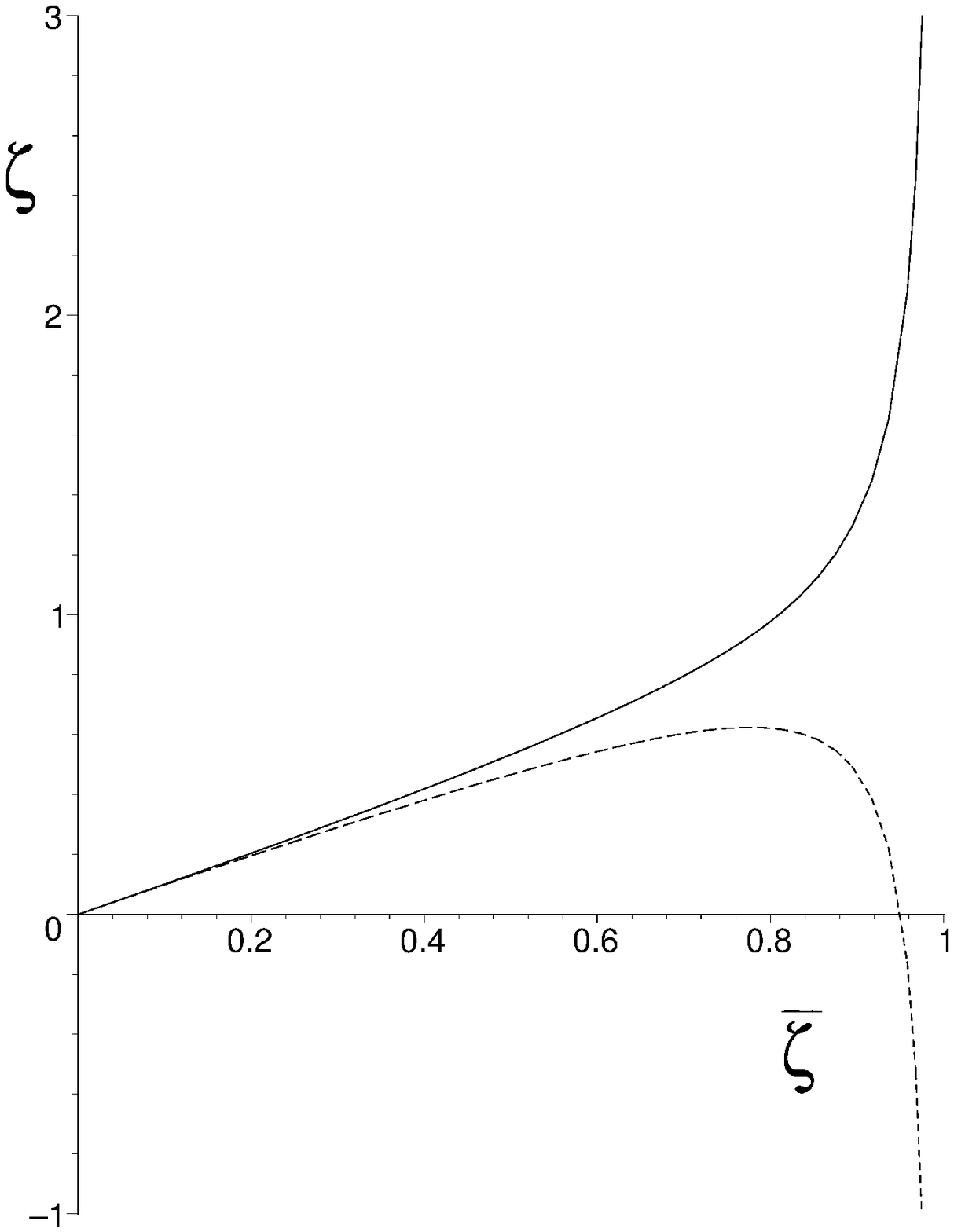}{3truein}

\fig{\sl Simple pole  perturbation,$ f(z) = \xi/z$, with $\xi=  1$ (full line) and
$\xi=-1$ (dashed line).}{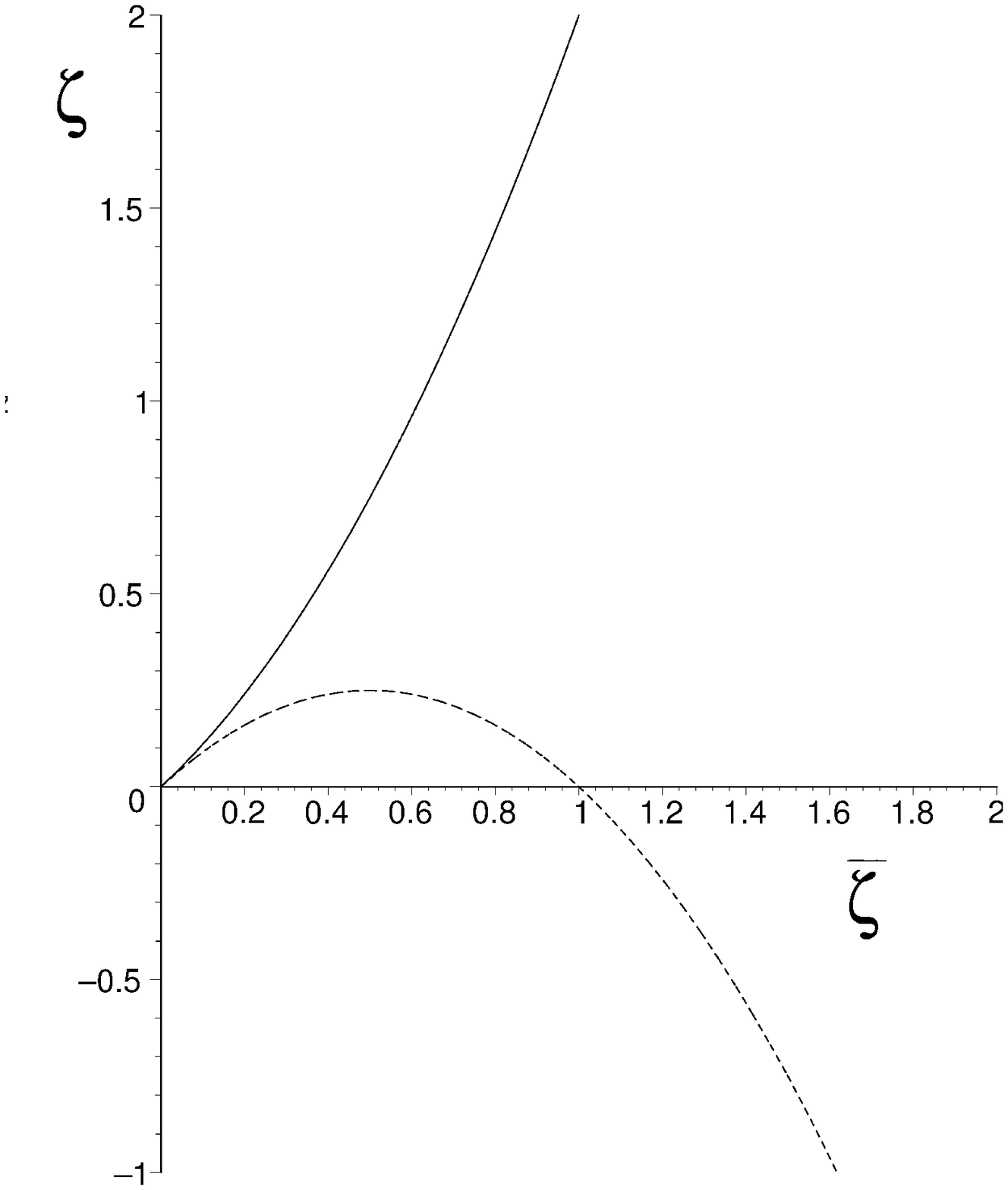}{3truein}

For small $\xi$ the behaviour is qualitatively equivalent to the
single-trace theory near the correspondence line $\ozeta =1$. However,
there are gross differences deep in the supergravity regime $\ozeta \ll 1$.

For $\xi >0$ we have well defined supergravity backgrounds with small
curvature ($\ozeta \rightarrow 0$) and {\it small} and negative bare
't Hooft coupling ($\zeta \rightarrow -\infty$). Restricting ourselves to
positive bare 't Hooft couplings we have a minimum value of $\ozeta$
given by
$\ozeta_c = \xi^{1\over 2n+1}$. At this point the bare 't Hooft coupling
diverges, i.e. $\zeta = H(\,\ozeta_c \,) =0$.

On the other hand, for $\xi <0$ the $H$-function has a minimum at
$\ozeta_m = (2n|\xi|)^{1\over 2n+1}$ at which the bare 't Hooft
coupling becomes maximal.  This is a critical point of the master
equation with the onset of a local instability for lower $\ozeta$. 
  For $\zeta > \zeta_m$ there are two
possible solutions of the master equation, but only the more  
curved one (larger $\ozeta$) is locally stable.    

For either sign of $\xi$ the dynamics differs significantly from the
single-trace model when the inverse 't Hooft coupling reaches values
of order $\zeta \sim |\xi|^{1\over 2n+1}$.  Hence, for $|\xi| \sim 1$
the single-trace behaviour is pushed beyond the correspondence line.
In particular, for $\xi \ll -1$ the supergravity regime loses all
stable solutions.

\subsec{Binomium Perturbation}

\noindent

For a  perturbation of the form
\eqn\binom{ 
f(z)=\xi \; {z^{n+1}\over n+1} +\eta \; 
{z^{l+1} \over l+1} \;, \;\;\;\;n>l\;,
}
one has
\eqn\hbin{
H(\,\ozeta\,) = \ozeta - {\xi \over \ozeta^{\;2n}} - {\eta \over
\ozeta^{\;2l}}
\;.}

  There is no qualitative difference with the
monomium if $\xi $ and $\eta $ have both the same
sign. If the sign is different and  $\xi <0$, the shape is the
same as in the monomium but the minimum is displaced to lower values of
$\zeta$, and can eventually  cross the $\zeta=0$ axis, removing  
the retrictions on the value of the 't Hooft parameter (c.f. Fig 3). 
The amount of the displacement is related to the ratio
$|\eta/\xi|$, growing with it up to a maximum value. If  $\xi
>0$ a maximum and a minimum of $H(\,\ozeta\,)$ appear if the
 ratio $|\eta/\xi|$ is large enough.  
 Thus, a new unstable branch
develops and there is a phase transition between the low-$\ozeta$
branch and the high-$\ozeta$ branch as $\zeta$ increases (c.f. Fig 4).

\subsec{More Exotic Perturbations}

\noindent

In the supergravity approximation one
 may also study non-polynomial ``perturbations" at a formal level.
For example, we can consider an analytic   
 logarithmic perturbation $f(z)=\xi\,\log\,(\varepsilon z -1)$, with $0< \varepsilon <1$,
 leading to
\eqn\hlog{
H(\,\ozeta\,) = \ozeta - \xi\,\varepsilon\;{ \ozeta^{\;2}  \over  \varepsilon -\ozeta^{\;2}}\;.}
The function $f(\,\ozeta\,)$ develops an imaginary part for $\ozeta > \sqrt{\varepsilon}$,
so that we are led to the restriction $\ozeta < \sqrt{\varepsilon} <1$. If $\varepsilon >1$
this boundary goes beyond the supergravity regime.  The $\xi<0$ branch is locally
stable, whereas the $\xi>0$ branch is only stable up to a maximum value of $\ozeta$.
This determines in turn a maximum value of $\zeta$ (or a minimum 't Hooft coupling).

An even more formal perturbation is given by the non-analytic  
function  $f(z)=\xi/z$ with 
\eqn\hpolo{
H(\,\ozeta\,) = \ozeta + \xi \; \ozeta^{\;2}\;. 
}
In this case, the perturbation by $1/\tr F^2$ does not make much sense in
perturbation theory. However, the results are quite smooth in the supergravity
approximation, corresponding to the limit of very large 
't Hooft coupling. Indeed, from \hpolo\ we see that these models produce
an {\it analytic} function $H(\,\ozeta\,)$, so that they  
approach the single-trace theory for $\zeta \ll 1$.  
 If $\xi<0$ there is a maximum of the master
equation, which means that the 't Hooft parameter should be higher
than $4\,|\xi|$.

In general, we see that novel qualitative  features triggered by 
multitrace couplings  stay well within  the supergravity approximation
only as long as $|\xi | \ll 1$, with $\xi$ a generic multitrace coupling.   
Since the master equation takes the form
\eqn\mmma{
\zeta = H(\,\ozeta\,) = \ozeta -f'\left(\,\ozeta^{\,-2}\,\right)\;,}
we find that deformations that have $f'(z)$ analytic around the origin
produce important qualitative changes in the deep supergravity regime
$\ozeta \rightarrow 0$. Conversely, singular deformations in perturbation
theory, corresponding to singular $f'(z)$ at the origin, approach the  
single-trace theory in the extreme supergravity regime.  

\newsec{Multitraces and Topology-Changing  Phase Transitions}

\noindent 

 Large-$N$ phase transitions induced by multitrace couplings arise
from the many possible solutions of the master equation $\zeta = H(\,
\ozeta_i\,)$. However, in most cases considered so far the different
solutions $\ozeta_i$ are continuously connected and the corresponding
string backgrounds have the same topology when studied in the supergravity
approximation. 
Large-$N$ phase transitions with change of spacetime topology are known
in the AdS/CFT framework, the most famous example being the Hawking--Page
transition \refs{\rHPage, \rwitthp}, corresponding to a CFT on a 
finite-radius sphere. From the CFT point of view, the phase transition
arises as a finite-size effect. 

The interplay between multitrace-induced and topology-changing transitions
is a interesting question that we address in this section. Fortunately,
the example  model  based on hot D-branes does show topology-changing
transitions when the Yang--Mills theory is compactified on a torus, so that
we can carry on our study in a rather direct way.
We start with a short review of the topology-changing transitions
corresponding to finite-size effects  of SYM models  
on toroidal compactifications.

\subsec{Review of the Single-Trace Case}

\noindent

Let us consider the compactification of the hot D4-brane on a
 $(4-p)$-dimensional  torus
of size $L$. In the perturbative description,  
 the Euclidean spacetime of the SYM model
 at finite temperature has the topology
${\bf R}^p \times {\bf S}^1_\beta \times \left({\bf S}_L^1 \right)^{4-p}$, with
$\beta = 1/T$ and we take the supersymmetric spin structure on the torus of
size $L$.    

In the perturbative regime, $\lambda \ll 1$, the thermodynamics of the
${\rm SYM}_{4+1}$ theory  changes character at
 $TL \sim 1$, from five-dimensional
scaling of the entropy $S\sim T^4$ at $TL >1$ to a $p$-dimensional        
scaling $S\sim T^{p-1}$ at $TL <1$.  
At strong coupling $\lambda \gg 1$, the AdS/CFT correspondence incorporates
this change of behaviour by a transition between topologically distinct
backgrounds, both with the same asymptotic boundary conditions 
\refs\rthresholds. The
first background is the near-horizon geometry of the original
black D4-brane wrapped on the $(4-p)$-torus.   This metric is T-dual
to that of black D$p$-branes, localized on the torus, but
 distributed uniformly over its volume, i.e. the so-called  black brane
``smeared" over $4-p$ transverse dimensions. 

The vacuum energy per unit volume in ${\bf R}^p$ is the same for both
T-dual metrics and is given by the thermodynamic free energy of the
five-dimensional theory, i.e. we can write 
\eqn\oos{
w_0 = -{1\over N^2\, V_p}\;\log\;\CZ_0 (T)\;,}
where $V_p$ is the volume in the noncompact ${\bf R}^p$ directions. 
 In general, for a $(d+1)$-dimensional SYM theory
at finite temperature $T$ on a spatial volume $V_d$ we have (c.f.
\refs{\rmaldacobi,\rthresholds})  
\eqn\freen{
-\log\,\CZ_0 (T) = -N^2 \;V_d\;C_d\;
(g^2_{\rm eff} N)^{d-3 \over 5-d} \;T^{\;{9-d \over
5-d}}\;,}
where
$g_{\rm eff}^2$ is the effective SYM coupling constant  of mass dimension
$3-d$.  
Considering now the particular case of D4-branes wrapped on the
${\bf T}^{4-p}$ torus, the effective coupling is  
$g_{\rm eff}^2 N = \lambda /T$ and
\eqn\vaeen{
w_{0,s} = -L^{4-p}\, C_4\, \lambda\;T^{\;4}\;.}

With the same quantum numbers and asymptotic behaviour,
 one can consider the metric of D$p$-branes
 fully localized on the $(4-p)$-torus. The corresponding vacuum 
energy is  related to the thermodynamic free energy of the effective
 SYM theory in $p$ Euclidean dimensions, with effective coupling
$g_{\rm eff}^2 N = \lambda\,L^{\;p-4} /T$,   
\eqn\valoc{
w_{0,\ell} = -N^2 \,C_p \, \left(\lambda L^{p-4} /T \right)^{p-3 
\over 5-p} \;T^{9-p \over 5-p}\;.}

The smeared metric dominates for large temperatures, whereas the
localized metric takes over at low temperatures. The cross-over
temperature is given by 
\eqn\ratio
{ 1= {w_{0,s} \over w_{0,\ell}} = {C_4 \over C_p}\;
\left(\,\lambda \,L\,T \,\right)^{2(p-4) 
\over p-5} \;,}    
which defines the ``localization curve" 
\eqn\curvl{
\lambda \sim  {1\over L\,T} \;.}

The transition between the smeared and the localized geometry is
reminiscent of the Gregory--Laflamme instability \refs\rgl. They are, however,
very different, since both backgrounds are locally stable in the
near-horizon regime (they both have positive specfic heat \refs\rgm). Thus, we
have a first-order phase transition between locally stable backgrounds. 

\fig{\sl Phase diagram of the finite-temperature D4-brane at large $N$,
 as a function
of the effective dimensionless coupling  $\zeta = 1/\lambda$  and the
size of the torus $LT$, in units of the temperature. Full lines denote
finite-size localization transitions and dashed lines correspond
to correspondence regions between supergravity and perturbative
descriptions.}
{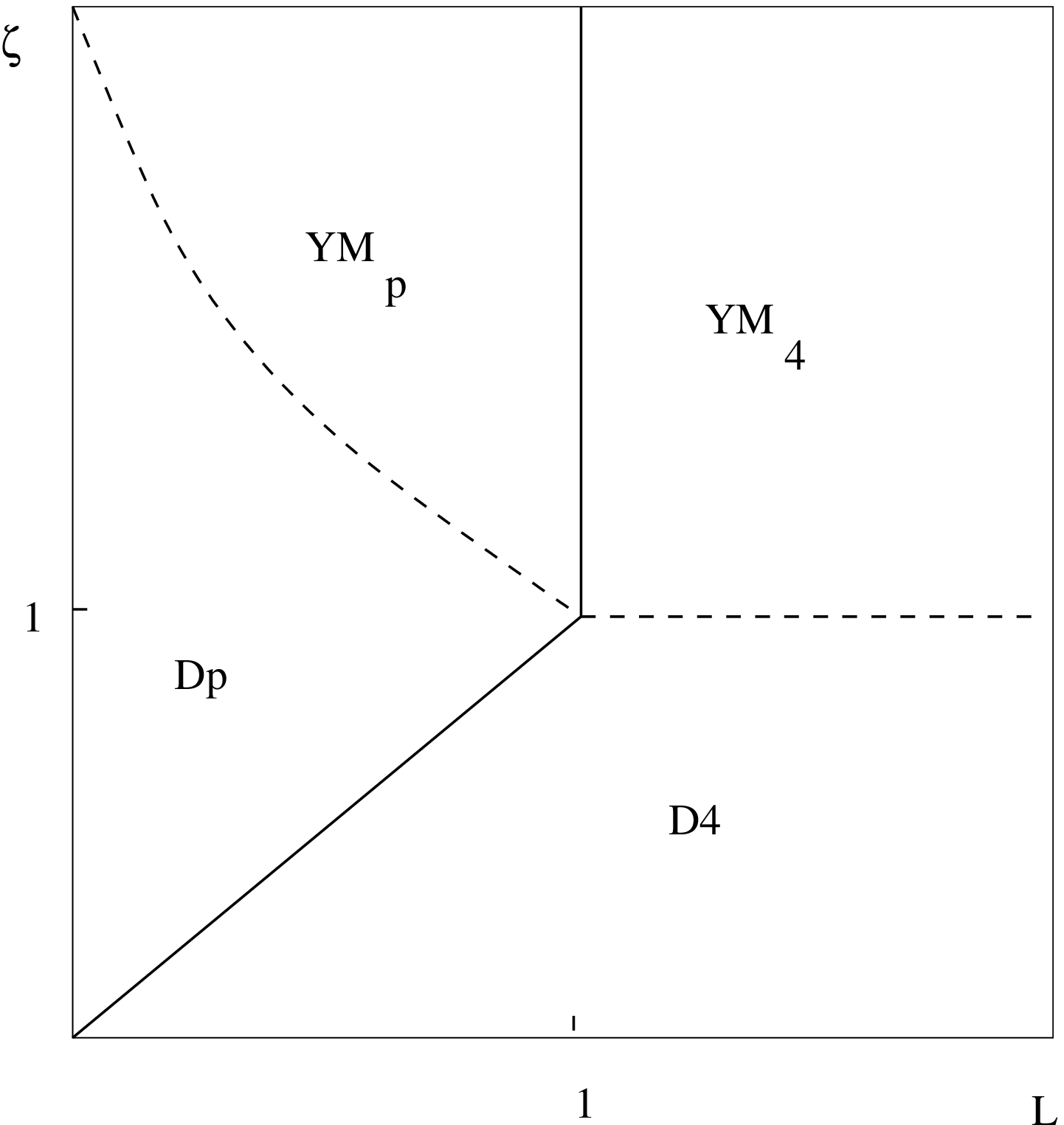}{3truein}

We plot in Fig 7 a phase diagram of the single-trace theory as
a function of $\zeta = 1/\lambda$ and the dimensionless combination $LT$. 
The localization curve \curvl\  continues at weak coupling as $LT=1$.
The correspondence lines separating the perturbative and the supergravity
regimes are $\zeta =1$ for large $LT$ and $\zeta = (LT)^{p-4}$ for
smaller values of $LT$.

\subsec{The Multitrace Case}

\noindent

Since the multitrace deformation modifies de vacuum energy through
\ledd\  the phase transition curves change accordingly.
Let us consider the simple case of a monomium perturbation
of the form \spp\ in the $d=4$ case. 
 The ``smeared" phase over the $(4-p)$-torus   
of size $L$ has vacuum energy density
\eqn\smend{
w_s = - L^{4-p}\,\ozeta_s^{\,-1} -  L^{4-p}\,
\xi \;{n \over n+1} \,\left(\ozeta_s^{\,-2}
\right)^{n+1}\;,}
where we have chosen couplings so that $C_4=1$ and we use units with $T=1$
throughout this section.  $\ozeta_s$ denotes
the selfconsistent  coupling  in the smeared phase, obeying the master
equation
\eqn\mse{
\zeta = \ozeta_s -\xi\,\ozeta_s^{\,-2n}\;.}

On the other hand, the phase of localized black D$p$-branes
yields similar expressions after dimensional reduction to ${\bf R}^p$. 
The effective dimensionless coupling arising 
 through the standard  rule
\eqn\ruled{
{1\over \lambda} \int d^4 x \longrightarrow {L^{\,4-p} \over \lambda} 
\int d^p x}
is given by
\eqn\effdm{
\zeta_{\rm eff} = L^{\,4-p}\,\zeta\;,}
and similarly for $\ozeta$. The master equation of the localized phase
is then 
\eqn\locmas{
\zeta = \ozeta_\ell - \xi\,\left[(p-3)\,C_p\,\left(\,\ozeta_\ell \,L^{\,4-p}
\right)^{2\over p-5} \right]^n\;.}
The resulting vacuum energy is given by 
\eqn\valoc{
w_\ell = -(5-p)\,C_p \,\left(\,
L^{\,4-p} \;\ozeta_\ell \,\right)^{3-p \over 5-p} - \xi\,{n \over n+1}
\,\left[ (p-3)\,C_p\,\left(\,\ozeta_\ell \,L^{\,4-p}\right)^{{2\over p-5}} \right]^{n+1}
\;.}

\fig{\sl Phase diagram for the localization transition between  D4-branes and
D3-branes with a multitrace perturbation $\xi<0$. The localization line
terminates at $(\zeta_m, L_m)$ and the dotted line signals the local instability
of the D4-branes, induced by the multitrace couplings.}
{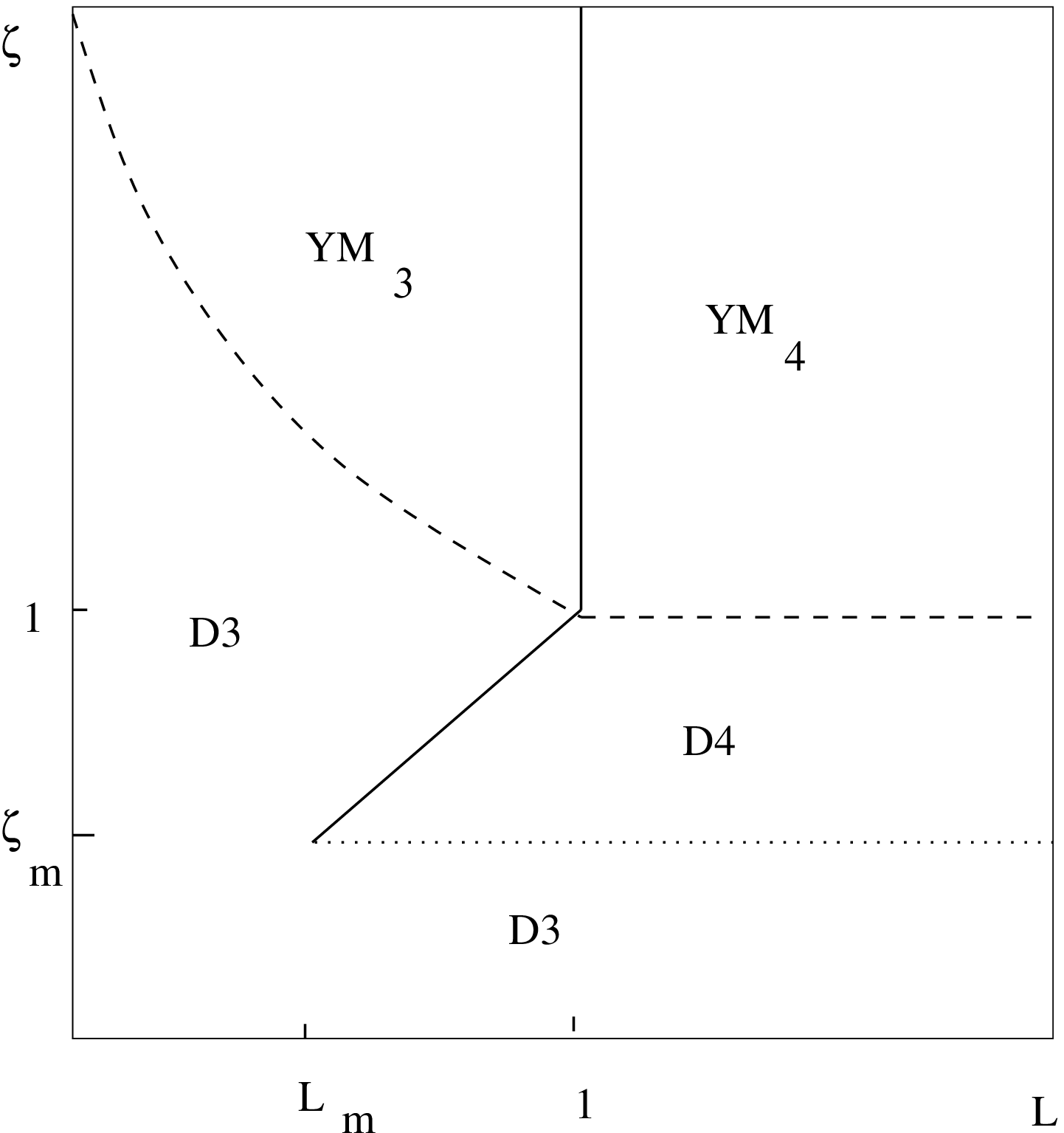}{3truein}

In general, we see that the multitrace perturbation  is not qualitatively
significant for
\eqn\limi{
\zeta \gg |\xi|^{1\over 2n+1}
}
   in the smeared phase. In the localized phase, the qualitative features
are standard for
\eqn\limm{
\zeta \gg L^{\,p-4}\,|\xi|^{5-p \over 5-p + 2n}
\;.}
We shall further simplify the analysis by choosing $p=3$, i.e. compactification
on a single circle of size $L$. In this particular case the equations \locmas\
and \valoc\ governing the localized phase collapse to very simple expressions:
\eqn\ptres{
\zeta = \ozeta_\ell \;,\qquad w_\ell = -C_3\;.}
The transition curve between the smeared and the localized phases in the supergravity
regime 
 follows from the equation
\eqn\ttc{
1= {w_s \over w_\ell} = {L \over 2C_3 \,\ozeta_s} \left[ 1+ {n \over n+1}
{\xi \over \ozeta_s^{\,2n+1}} \right]\;.}

For $\xi <0$ we have a minimal value of $\ozeta_s$ for which the smeared solution
is locally stable:
\eqn\locmm{
\ozeta_m = \left(2n\,|\xi|\right)^{1\over 2n+1}\;.}
The main effect of this in the $(\zeta, L)$ phase diagram is the abrupt termination
of  the localization line
\ttc\  at the point $(\zeta_m, L_m)$,  with
\eqn\termi{
\zeta_m = {2n+1 \over 2n} \,\left(2n\,|\xi|\right)^{1\over 2n+1}\;, \qquad L_m = 
{2\,C_3\,(n+1) \over 2n+1} \;\left(2n\,|\xi|\right)^{1\over 2n+1}\;.}
Hence, part of the supergravity regime that was dominated by D4-branes is now
covered by the D3-brane phase due to the local instabilities induced by the
multitrace coupling (see Fig 8).

\fig{\sl Phase diagram for the localization transition between  D4-branes and
D3-branes with a multitrace perturbation $\xi>0$. The localization line
intersects the $\zeta=0$ axis at $L=L_c$, cutting off part of the D4-brane phase.
In this case, there are no local instabilities.}{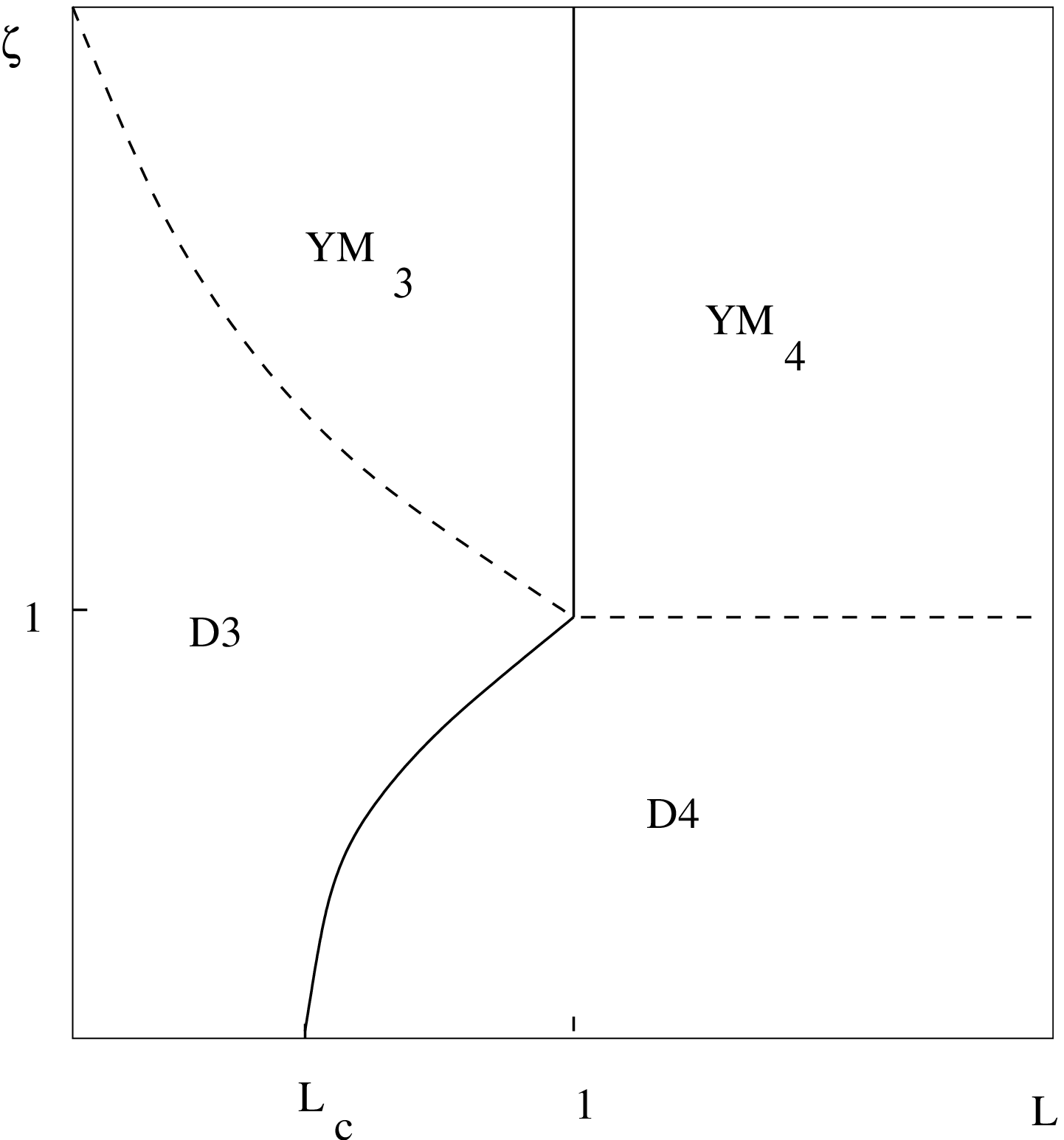}{3truein}

For $\xi>0$ the D4-brane phase is locally stable, but the condition $\zeta >0$
still enforces a minimum value of $\ozeta_s$, given by 
$\ozeta_c^{\,2n+1} = \xi$. The critical length at this point is
\eqn\clengg{ 
L_c={n+1\over 2n+1}\;2\,C_3 \,\ozeta_{c}
\;.}
Expanding the master equation  \mse\ and the localization equation \ttc\ near
the point $(\zeta=0, L_c)$ we find 
\eqn\appri{
\zeta \approx {2n+1 \over n+1}\;\ozeta_c\;\left(1-{L_c \over L}\right)}
in the vicinity of the $\zeta=0$ axis. The resulting phase diagram is depicted in
Fig 9. 

Thus, comparing with Fig 7, we see that the multitraces tend to modify the
structure of topology-changing phases in the extreme supergravity regime (low values
of the effective coupling $\ozeta\;$). 

\newsec{Conclusions}

\noindent

We have studied multitrace perturbations of large $N$ master fields in
theories with adjoint fields and non-vanishing mass gap. The introduction of suitable
auxiliary fields allows us to discuss the systematics of the $1/N$ expansion
at a formal level. At the same time, we were able to derive the ``master
equation" of the Hartree approximation \meq, and define criteria for local
stability of the master fields.

The example of hot D4-branes was studied in some detail. In particular, we
find the conditions for the model to develop tachyons induced by multitrace
operators. The qualitative features of many types of ``master equations",
including those of non-polynomial multitraces, were analyzed in the
supergravity approximation, as well as the effects of these deformations
on the thermodynamic phase diagrams at finite volume.  
In general, we find that the pattern of large $N$ phase transitions is
strongly affected by the multitraces when the perturbation is of $O(1)$
in dimensionless units.

The main result of this paper is the relation between the global properties
of the master equation, in particular the monotonicity of the function $H(\,\ozeta\,)$
in \defh, and
the local stability of the master field. Since multitrace perturbations in the AdS/CFT
context correspond to the  nonlocal string theories of \refs\rexodef, we have found
new classes of tachyonic instabilities of these models, entirely induced by the
multitrace couplings. These instabilities arise despite the fact that,
 within the Hartree approximation, one
can compute operator condensates in an auxiliary 
single-trace model (the background ${\overline X\;}$).
  It would be interesting to sharpen the connection between
these instabilities and the so called ``stringy exclusion principle" of 
\refs\rexpri.   

Further lines of research suggested by our work include a more precise 
interpretation of the tachyons in the ``bulk" spacetime description (as
peculiar instabilities of ${\overline X}\;$) and the relation to
supersymmetry breaking (in our examples, supersymmetry is broken to generate
a mass gap, already at the level of the single-trace model).  In particular,
some type of  UV/IR dictionary for the composite field $\sigma(x)$ would be 
required to understand how these instabilities proceed in the bulk.  
It should also be interesting to pursue the calculation of $1/N$ effects,
starting with the one-loop determinants, for example along the lines of
\refs\rgubmit.

\vskip1cm

{\bf Acknowledgements}

We would like to thank Enrique \'Alvarez for many discussions, suggestions and
collaboration in the initial stages of this work. The work of J.L.F.B. 
was partially supported by MCyT
 and FEDER under grant
BFM2002-03881 and
 the European RTN network
 HPRN-CT-2002-00325. The work of C.H. was partially supported by 
European Commission (HPRN-CT-200-00148) and CICYT (Spain) and 
 by the MECD (Spain) through a FPU grant.

\appendix{1}{Topological Charge and Multitrace Perturbations}

\noindent

In the general discussion of Section 2 we assumed the standard large $N$
scaling of one-point functions $\lim_{N\to \infty} \bra \CO \ket = O(1)$,
with the normalization $\CO \sim N^{-1} \;\tr \,F^n$. The purpose of this
appendix is to point out an important exception of this rule, namely the
case of the topological operator
\eqn\topol{
\CQ = {1\over 8\pi^2} \,\tr \,F\wedge F\;.}
Notice the absence of the $N^{-1}$ factor in front. Despite this
enhanced normalization, the large $N$
one-point function $\bra \,\CQ\,\ket$  
is still of $O(1)$, 
 provided the $\theta$ angle is nonvanishing in the action     
\eqn\acc{
S_0={1\over 2g^2}\int d^4 x\;
 \tr \,F^2+{i\theta\over 8\pi^2}\int \tr F\wedge F
\;.}

Extracting an explicit power of $N$ in the normalization of the topological
 action, we see that it is $\theta/N$ what plays the role of the
 't Hooft coupling. Naively this
 is incompatible with $2\pi$ periodicity in $\theta$, but it can be 
  restored if the effective action develops 
  multiple branches (c.f. \refs\rwittenold)  
\eqn\mbra{
\CZ(\theta)=\exp\,\left[-N^2 W_0 (\theta/N)\right]=\int D A \;e^{-S_0}
 =\sum_k \exp\left[-N^2 \,V\,h\left( {\theta+ 2\pi k\over N}\right)\right]
\;,}
where $V$ is the Euclidean spacetime volume,   
  and the function $h(y)$  
is assumed to have 
 a Taylor expansion with $O(1)$ coefficients  in the large $N$ limit. 
 
For each $\theta$, the sum over $k$
 is dominated by the least-action branch in the
infinite volume limit, 
\eqn\lactb{
W_0 (\theta/N)\approx  {\rm min}_{\;k} 
\;V\,h\left( {\theta+ 2\pi k\over N}\right)
\;.}
The solution of the $\eta'$ puzzle by Witten and Veneziano \refs\retap\
 requires that the function $h(y)$ behaves as $h(y)\sim y^2$ near the origin. 
More specifically, let us assume the result of Ref. \refs\rwitheta\
 in the AdS/CFT case and set
\eqn\hwi{
h(y)=\shalf\,C\;y^2+ \CO(1/N^2)
\;.}
It follows that the full  theta-dependence of the vacuum
energy is also of  $O(1)$.  For example,  
  the one-point function of $\CQ$ scales as 
\eqn\opfQ{
\bra\,\CQ\,\ket={i\over V}\partial_\theta \,\log\,
 \CZ(\theta) = -i N^2 {1\over N}
 h'\left({\theta + 2\pi k_c \over N}\right)\approx -i \,C\, 
(\theta + 2\pi k_c)+\CO(1/N^2)
\;,}
where $k_c$ is the integer that minimizes \lactb. 
Therefore,  the one-point function of the topological charge is  of
$O(1)$ in the large $N$ limit, despite its anomalous  
anomalous normalization as it appears in  \topol. 

The connected two-point function is related to the topological susceptibility
as 
\eqn\topsus{
{\overline Q_{\rm top}^2}=\int_x \bra \CQ(x) \CQ(0) \ket_c={1\over V} 
\int_x\int_y \bra \CQ(x) \CQ(y)\ket_c=-{1\over V} \;\partial_\theta^2 \,
\log \,\CZ(\theta)\;, 
}
and this in turn determines the constant $C$ in \hwi,  
\eqn\topsusb{
{\overline Q_{\rm top}^2}= {1\over V} N^2{1\over N^2} h''\left( {\theta+2\pi k\over N}\right)\approx C\;.
}

The peculiar scaling properties of the topological
 charge operator affect the mean field analysis of the multitrace perturbations. Consider the perturbed action
\eqn\multopac{
S=S_0+\int_{{\bf R}^4} f\left(\CQ\right)
\;.}
Introducing the auxiliary fields and taking into
 account the assumed  multibranched structure of the single-trace theory
we find 
\eqn\parfmult{
\CZ(\theta)=\sum_k \int D\,\sigma\;D\,\chi\;
 \exp\left[ -N^2 \Gamma\left({\theta +\chi +2\pi k\over N}\right)-
\int f(\sigma)+i\int\chi\,\sigma\right]
\;,}
where the functional $\Gamma/V$ equals the function $h$ in \mbra\ when
evaluated on constant functions. 

 There is a saddle point for each of the branches, satisfying the equations  
\eqn\sddla{
\chi_k=f'(\sigma_k)
\;, \qquad 
\sigma_k=\bra\, \CQ\, \ket_{\otheta_k}
\;,}
where we have defined the effective theta angle 
$\otheta_k\equiv \theta+\chi_k+2\pi k
$, so that the  master equation for each branch reads  
\eqn\meqtop{
\theta+2\pi k=\otheta_k-f'\left(\,\bra \,\CQ \,\ket_{\otheta_k}\right)
\;.}
The quadratic fluctuation operator is 
\eqn\wrons{
\CK_\CQ=\pmatrix{\bra \,\CQ \,\CQ\,\ket_{c,k} & -i \cr -i &
 f''(\,\bra \,\CQ\,\ket_k )}
\;,}
and its contribution to the vacuum energy in each branch is
\eqn\contb{
 \shalf \;\Tr\;\log\;\left[1+ f''(\sigma_{k})
 \; \bra\,\CQ \CQ\,\ket_{c,k} 
\right] \;.}
It contributes to the theta-dependence at the   $O(1)$ order, just
as the function $h(y)$ above. Hence, one must calculate the determinant
in order to find the balance between the different branches.

\vskip0.2cm

\listrefs

\bye